\documentclass[twocolumn]{IEEEtrans}

\def\BibTeX{{\rm B\kern-.05em{\sc i\kern-.025em b}\kern-.08em
    T\kern-.1667em\lower.7ex\hbox{E}\kern-.125emX}}

\begin{document}

\title{A Bayesian Estimator for Linear Calibration Error Effects in Thermal Remote Sensing}

\author{J. A. Morgan\\The Aerospace Corporation\\P. O. Box 92957\\Los Angeles, CA 90009}



\maketitle

\begin{abstract}

The Bayesian Land Surface Temperature estimator previously developed has been extended to include 
the effects of imperfectly known gain and offset calibration errors.  It is possible to treat both 
gain and offset as nuisance parameters and, by integrating over an uninformitave range for their 
magnitudes, eliminate the dependence of surface temperature and emissivity estimates upon the exact
calibration error.

\end{abstract}


\section{Introduction}

\PARstart{A}{s} a practical matter, land surface temperature (LST) estimates retrieved from radiances 
reported by a remote sensor will 
be subject to some unavoidable level of calibration error, which need not be very accurately known.  
While all forms of quantitative exploitation of radiometric data in remote sensing are afflicted to 
some degree by calibration error effects, this problem may be a special concern for the Bayesian 
multiband LST algorithm~\cite{Morgan2001}.  That is because the Bayesian algorithm iterates on a range 
of plausible surface temperatures, within which the estimated LST value is obtained as an expected 
value.  Should uncompensated calibration errors lead to a temperature interval which does not bracket 
the true surface temperature, the algorithm in its present form has no way to recover, and may return 
a surface temperature estimate with degraded accuracy.   

This note sketches the extension of the Bayesian approach to LST retrieval to include effects of 
a simple form of unknown calibration error.  After a review of the Bayesian approach to LST
retrieval, the calibration error is parameterized as linear in the true aperture radiance.  The joint 
prior probalility for the calibration error parameters is then obtained by imposing 
the requirement that two distinct observers agree on its mathematical form.  Finally, the Bayesian 
LST estimator is extended to include linear calibration error by treating the calibration error 
parameters as nuisance variables, and integrating them out of the final estimators for surface 
temperature and emissivity.    

\section{Elements of Bayesian LST estimator}

The Bayesian land surface temperature retrieval algorithm is developed in~\cite{Morgan2001}, which may be
consulted for details.  The approach to LST retrieval presented in that earlier paper consists of three 
elements:

1.  The forward model for sensor aperture radiance, assumed linear in surface emissivity:
\begin{equation}
I_{F}(k)=\epsilon_{k}B_{k}(T)exp(-\frac{\tau_{k}}{\mu})+
\frac{\rho_{k}}{\pi} F_{k}^{\downarrow}
(0)exp(-\frac{\tau_{k}}{\mu})+I_{k}^{\uparrow}(\tau,\mu) \label{eqn2}
\end{equation}
$I_{k}^{\uparrow}(\tau,\mu)$ and $F_{k}^{\downarrow}(0)$ are the upwelling diffuse radiance at nadir 
optical 
depth $\tau$ (top of the atmosphere, or TOA, for spaceborne sensors; $\mu$ is the cosine of the angle 
with respect to zenith) and the downwelling irradiance at the surface, respectively.  $B_{k}(T)$ is the 
Planck function at surface temperature $T$.  The emissivity is $\epsilon_{k}$, and the surface 
reflectance $\rho_{k}=1-\epsilon_{k}$.  Note that (\ref{eqn2}) assumes Khirchoff's law; this is done 
solely for simplicity.  It is also assumed (at least initially) that the sensor has high spectral 
resolution.  

2.  The MAXENT form of the conditional probability of observing radiance 
$I$~\cite{Bretthorst1987},\cite{Bretthorst1988},\cite{Bretthorst1988a} in the presence of
noise-equivalent radiance $\sigma$:
\begin{equation}
P(I\mid T,\epsilon_{k},\sigma)=
exp\left[-\frac{(I-I_{F})^{2}}{2\sigma^{2}}\right]\frac{dI}{\sigma} \label{eqn7}
\end{equation}

3.  The prior probability of surface temperature and emissivity~\cite{Morgan2001}:
\begin{equation}
P(T,\epsilon_{k}\mid K)=\frac{\mbox{const.}}{T}dTd\epsilon_{k} \label{eqn45}
\end{equation} 

The posterior probability for the surface temperature and emissivity, given observed radiance and 
available knowledge, is obtained from these quantities by use of Bayes' theorem:
\begin{equation}
P(T,\epsilon_{k}\mid I,K) \propto
P(T,\epsilon_{k}\mid K) P(I\mid T,\epsilon_{k},\sigma). \label{eqn3}
\end{equation}

\section{Calibration error model}

By hypothesis, the physical radiance $I_{0}$ at wavenumber $k$ is related linearly to the reported 
radiance $I$:
\begin{equation}
I_{0}=(1+\alpha) I +\beta  \label{radredef}
\end{equation}
Both $\alpha$ and $\beta$ are assumed to be small quantities; 
\begin{equation}
\alpha << 1
\end{equation}
and
\begin{equation}
\beta << I_{0}
\end{equation}
It is the physical radiance $I_{0}$ which goes into (\ref{eqn2}).  Recalling that the forward model 
(\ref{eqn2}) is linear in $\epsilon$, the exponent in that expression is
\begin{equation}
(I_{0}-I_{F})^2 = (((1+\alpha) I +\beta)-(A \epsilon + B))^2 
\end{equation}
which is quadratic in $\alpha$, $\beta$, and $\epsilon$.

\section{Prior probability for calibration error parameters}

In order to obtain a useable estimator, it is necessary to find the prior probability for the 
calibration error parameters $\alpha$ and $\beta$~\cite{Jaynes1968},\cite{Jaynes1973},\cite{Jaynes1980}.
As in~\cite{Morgan2001}, two equally cogent 
observers must relate their descriptions of radiance, and of calibration error, by a Lorentz 
transformation~\cite{MTW1973},\cite{Weinberg1972} connecting one (primed) coordinate description 
moving with velocity 
$v$ along the observation axis with respect to the other (unprimed) one by
\begin{equation}
k'=\eta k \label{eqn19}
\end{equation}
where the Doppler factor $\eta$ is given in terms of the boost parameter
\begin{equation}
\gamma=\frac{1}{\sqrt{1-(v/c)^{2}}}. \label{eqn21}
\end{equation}
by 
\begin{eqnarray}
\eta \equiv \gamma (1-v/c) \nonumber \\
= \sqrt{\frac{1-v/c}{1+v/c}} \label{eqn20}
\end{eqnarray}
The quantity $\eta$ is real and nonvanishing for physical Lorentz transformations.  
Let 
\begin{equation}
P(\alpha,\beta \mid K)=g(\alpha,\beta)d \alpha d\beta \label{eqn20a}
\end{equation}
be the prior probability assigned by Vladimir in the unprimed frame, and
\begin{equation}
P(\alpha',\beta' \mid K)=h(\alpha',\beta')d \alpha' d\beta' \label{eqn20b}
\end{equation}
be that assigned by Estragon, viewing radiance in the primed frame.
The prior probabilities 
in the two frames are related by
\begin{equation}
h(\alpha',\beta')d \alpha'd \beta'=J^{-1}g(\alpha,\beta)d \alpha d\beta \label{eqn13}
\end{equation}
where
\begin{equation}
J=det\left[\frac{\partial{(\alpha',\beta')}}{\partial{(\alpha,\beta)}}\right] \label{eqn14}
\end{equation}
is the Jacobian determinant for the transformation.

Consider first $\alpha$ as defined by Vladimir.  Suppose that $\beta=0$; then by the Lorentz 
invariance properties of spectral radiance~\cite{MTW1973a} we have
\begin{equation}
\frac{I_{0}(k)}{k^{3}}=\mbox{invariant}=\frac{(1+\alpha) I(k)}{k^{3}}
\end{equation}
and also
\begin{equation}
\frac{I(k)}{k^{3}}=\mbox{invariant}
\end{equation}
if Vladimir and Estragon are to agree that the quantity $I(k)$ admits interpretation as a radiance.  
The ratio
\begin{equation}
\frac{(1+\alpha) I(k)}{I_{0}(k)}=\frac{\frac{(1+\alpha) I(k)}{k^3}}{\frac{I_{0}(k)}{k^3}}
\end{equation}
is likewise invariant, so that
\begin{equation}
1+\alpha=\mbox{invariant},
\end{equation}
as it must be, as the ratio of two radiances evaluated in the same Lorentz frame.  Thus
\begin{equation}
\alpha'(k')=\alpha(k)=\alpha(k')
\end{equation}

Next consider $\beta$.  We have 
\begin{equation}
\frac{(1+\alpha) I +\beta}{k^3}=\mbox{invariant}
\end{equation}
from which
\begin{equation}
\frac{\beta}{k^3}=\mbox{invariant}
\end{equation}
as must be for any radiance, in particular a noise radiance.  By (\ref{eqn19}) we find
\begin{equation}
\beta'(k')=\beta'(\eta k)=\eta^3 \beta(k)
\end{equation}

The Jacobian is therefore
\begin{equation}
det\left[\frac{\partial{(\alpha',\beta')}}{\partial{(\alpha,\beta)}}\right]=\eta^3
\end{equation}
so Vladimir and Estragon must agree that
\begin{equation}
g(\alpha,\beta)d\alpha d\beta=\eta^3 h(\alpha',\beta')d\alpha'd\beta'
\end{equation}
and, by the principle of indifference~\cite{Morgan2001},\cite{Jaynes1968}, that
\begin{equation}
g(\alpha,\beta))=\eta^3 g(\alpha,\eta^3 \beta))
\end{equation}
with solution
\begin{equation}
g(\alpha,\beta) d\alpha d\beta=\frac{\mbox{const.}}{\beta} d\beta d\alpha \label{prior}
\end{equation}

Application of Bayes' theorem (\ref{eqn3}), as in~\cite{Morgan2001}, immediately gives the result that
the joint posterior probability of $T$, $\epsilon$, $\alpha$, and $\beta$ is proportional to 
the product of (\ref{eqn7}), (\ref{eqn45}), and (\ref{prior}):
\begin{equation}
P(T,\epsilon,\alpha,\beta,\sigma \mid I,K ) \propto
exp\left[-\frac{(I-I_{F})^{2}}{2\sigma^{2}}\right] 
d\epsilon \frac{dT}{T} d\alpha \frac{d\beta}{\beta} \frac{dI}{\sigma}. \label{drumroll}
\end{equation}

\section{Extended LST estimator}

Estimators for  $T$ and $\epsilon$ may be constructed from (\ref{drumroll}) as expectation
values in exactly the same 
manner as in~\cite{Morgan2001}.  The treatment of spectral quantities integrated over a passband 
follows the equivalent discussion in~\cite{Morgan2001}.  In practice, (\ref{drumroll}) will be 
unaltered for band-integrated radiances.  

The calibration error parameters $\alpha$ and $\beta$ are 
treated as nuisance parameters:  One does not care what their actual values are, so long as they
lie between specified limits. One is therfore at liberty to integrate (\ref{drumroll}) over those 
limits and obtain estimators for
\begin{equation}
\langle T \rangle = \frac{\int_{T_{min}}^{T_{max}} TP(T \mid I_{i},\sigma)\frac{dT}{T}}
{\int_{T_{min}}^{T_{max}} P(T \mid I_{i},\sigma)\frac{dT}{T}} \label{eqn61}
\end{equation}
and
\begin{equation}
\langle \epsilon_{i} \rangle =
 \frac{\int_{\epsilon_{min}}^{\epsilon_{max}}\epsilon P(\langle T \rangle ,\epsilon
 \mid I_{i},\sigma)d\epsilon}
{\int_{\epsilon_{min}}^{\epsilon_{max}}P(\langle T \rangle ,\epsilon \mid I_{i},\sigma)d\epsilon} 
\label{eqn63}
\end{equation}
in terms of
\begin{eqnarray}
\lefteqn{P(T,\epsilon \mid I,\sigma)=} \nonumber \\
&  \int_{\alpha_{min}}^{\alpha_{max}}
\int_{\beta_{min}}^{\beta_{max}} P(I\mid T,\epsilon,\alpha,\beta,\sigma) 
  d\alpha \frac{d\beta}{\beta}
\end{eqnarray}
and
\begin{eqnarray}
\lefteqn{P(T \mid I,\sigma)=} \nonumber \\
& \int_{\epsilon_{min}}^{\epsilon_{max}} \int_{\alpha_{min}}^{\alpha_{max}}
\int_{\beta_{min}}^{\beta_{max}} P(I\mid T,\epsilon,\alpha,\beta,\sigma) 
d\epsilon  d\alpha \frac{d\beta}{\beta}.
\end{eqnarray}
In (\ref{eqn61}) and (\ref{eqn63}), $\langle T \rangle$ and $\langle \epsilon \rangle$ 
have no dependence on exactly
what the calibration error parameters $\alpha$ and $\beta$ were, for a given reported 
sensor aperture radiance.  

It does not appear feasible to integrate moments of (\ref{drumroll}) in closed form.
However, by integrating over $\epsilon$ 
first, it is possible to take advantage of the closed-form result for the LST posterior probability 
derived in~\cite{Morgan2001}:
\begin{equation}
P(T\mid I,\sigma)\propto \frac{1}{\sqrt{a}} exp \left [-\frac{\left [c-b^{2}/4a \right]}
{2\sigma^{2}} \right ] H(\epsilon_{max},\epsilon_{min}) \! \label{eqn58}
\end{equation}
where
\begin{eqnarray}
H(\epsilon_{max},\epsilon_{min})=
erf\left [\frac{\sqrt{a/2}(\epsilon_{max}+b/2a)}{\sigma} \right] \nonumber \\ 
-erf\left [\frac{\sqrt{a/2}(\epsilon_{min}+b/2a)}{\sigma} \right] \label{eqn59}
\end{eqnarray}
for each band i.  As in~\cite{Morgan2001}, 
\begin{equation}
a=\left [\int_{k_{1}}^{k_{2}}\left (B_{k}(T)-\frac{1}{\pi} F_{k}^\downarrow(0)
\right)exp(-\frac{\tau_{k}}{\mu})dk \right ]^{2}, \label{eqn53}
\end{equation}
\begin{equation}
 b=b_{1} b_{2} \label{eqn54}
\end{equation}
with
\begin{equation}
b_{1}=2\left [\int_{k_{1}}^{k_{2}}\left (B_{k}(T)-\frac{1}{\pi} F_{k}^\downarrow(0)
\right)exp(-\frac{\tau_{k}}{\mu})dk \right ] \label{eqn54a}
\end{equation}
\begin{equation}
b_{2}=\left[\int_{k_{1}}^{k_{2}}\left 
(\frac{1}{\pi} F_{k}^\downarrow(0)exp(-\frac{\tau_{k}}{\mu})+I_{k}^\uparrow(\tau,\mu) 
\right)dk-I_{i} \right ], \label{eqn54b}
\end{equation}
and
\begin{equation}
c=\left[\int_{k_{1}}^{k_{2}}\left 
(\frac{1}{\pi} F_{k}^\downarrow(0)exp(-\frac{\tau_{k}}{\mu})+
I_{k}^\uparrow(\tau,\mu) \right)dk-I_{i} \right ]^{2}
\label{eqn55}
\end{equation}
In (\ref{eqn58}), as in (\ref{drumroll}), the sensor radiance that appear in the quantites $a$, $b$, and
$c$ is related to the physical radiance by (\ref{radredef}).  

The remaining integration over the nuisance variables $\alpha$ and $\beta$ is now two-dimensional, and 
any integration over $T$ to form the expectation value $\langle T \rangle$ makes for a third quadrature,
for the full calculation.  This is potentially awkward for routine evaluation, but the computational 
burden can be alleviated in special cases:

1.  If one knows $\emph{a-priori}$ that one is operating in a regime dominated by either gain or offset
calibration error, the less important source of error may be ignored as a first approximation.

2.  In a vicarious calibration, the surface temperature may be accurately known.

3.   Once a value for $\langle T \rangle$ is obtained for one pixel in a dataset, expectation values
$\langle \alpha \rangle$ and $\langle \beta \rangle$ can be calculated and used in estimation of 
$\langle T \rangle$ for other pixels.  Should calibration error be slowly varying, estimates of 
$\langle \alpha \rangle$ and $\langle \beta \rangle$ obtained from one dataset could be used for 
subsequent ones, or as initial guesses for updated estimates of 
$\langle \alpha \rangle$ and $\langle \beta \rangle$.

\end{document}